\documentstyle[12pt,a4,axodraw]{article}

\begin{document}
\setcounter{page}{0}

%========================================================================%
\thispagestyle{empty}
\begin{flushright}
                                                   TIFR-HECR-99-02 \\
                                                   hep-ph/9904234 \\
\end{flushright} 
\vskip 25pT
\begin{center}
{\Large\bf  
Constraining Large Extra Dimensions Using Dilepton Data from the Tevatron 
Collider }
\vskip 25pT
{\sl  Ambreesh K. Gupta~\footnote{E-mail: ambr@tifr.res.in}, 
      Naba K. Mondal~\footnote{E-mail: nkm@tifr.res.in}, 
      Sreerup Raychaudhuri~\footnote{E-mail: sreerup@iris.hecr.tifr.res.in}
{\footnotesize '}
\footnote{ 
Address after May 1, 1999: \\ 
\hspace*{0.2in} Department of Physics, Indian Institute of Technology, 
Kanpur 208 016, India.}} \\
\vskip 5pT
{\footnotesize\rm
Department of High Energy Physics, Tata Institute of Fundamental Research, \\
                Homi Bhabha Road, Colaba, Mumbai 400 005, India.} \\
\vskip 60pT
{\bf Abstract} 
\end{center} 
\vskip 5pT
{\rm 
We use the invariant mass distribution of Drell-Yan dileptons as measured
by the CDF and D\O~Collaborations at the Fermilab Tevatron and make a 
careful analysis to constrain Kaluza-Klein models with large extra 
dimensions. The combined data from both collaborations lead to a 
conservative lower bound on the string scale $M_S$ of about 1 TeV at 95\% 
confidence level. 
}
\vskip 120pt 
\begin{flushleft}
                                                   April 1999 
\end{flushleft} 

%========================================================================%
\newpage 

Recently, the idea that gravity could become strong at scales of the order 
of a few TeV has attracted a great deal of attention\cite{BaDiNe}. 
This is made possible if we allow for large compactified dimensions at the 
TeV scale. While such ideas can be fitted in within the scheme of quantum field 
theories\cite{ADD1}, a more natural construction\cite{ADD2,Sundrum} involves 
string theories with all Standard Model (SM) fields living on a 
three-dimensional D-brane (or 3-brane) embedded in a space of ($4+d$) 
dimensions (bulk). Of course, the original suggestion that we live 
in a spacetime continuum with more than the three canonical spatial 
dimensions was made early in this century \cite{KK}, but these 
Kaluza-Klein (KK) theories, as they are called, have not been able to 
satisfactorily reproduce the observed mass spectrum. Such ideas, however, 
have always formed a basic ingredient of string 
theories\cite{GrScWi}. In fact, models having extra 
dimensions with compactification scales of the order of a few TeV have been 
proposed\cite{Antoniadis} from time to time in the literature with various 
motivations. However, it is the discovery of D-branes\cite{Polchinski} which 
has provided the rather venerable KK theories with a new lease of life over 
the past year.

In a nutshell, the ideas proposed by Arkani-Hamed, Dimopoulos and Dvali 
(ADD)\cite{ADD1} and by Antoniadis {\it et al.}\cite{ADD2} are as follows. 
They suggest --- as all KK theories do --- that spacetime 
consists of ($4+d$) dimensions. The extra (spatial) $d$ dimensions 
are compactified, typically on a $d$-dimensional torus $T^d$ with radius 
$R$ each way. Since gravity 
experiments have not really probed the sub-millimetre regime, it is
proposed that $R$ can be as large as $\sim 0.1 - 1$ mm, a very large value 
when compared with the
Planck length $\simeq 10^{-33}$ cm. Though the actual value of
Newton's constant $G_N^{(4+d)}$ in the bulk is of
the same order as the electroweak coupling, its value $G_N^{(4)}$ in 
the effective
4-dimensional space at length scales $\gg R$ is the extremely small 
one measured in gravity experiments. This is decribed by a
simple relation derived\cite{ADD3} from Gauss' Law,
$$ \bigg[ M_{Pl}^{(4)} \bigg]^2 \sim R^d 
   \bigg[ M_{Pl}^{(4+d)} \bigg]^{(d+2)} 
$$ 
where $M_{Pl} \sim 1/\sqrt{G_N}$ denotes the Planck mass. 
If $M_{Pl}^{(4+d)} \sim 1$ TeV,
then $R \sim 10^{30/d - 19}$ m. This means that for $d = 1$, 
$R \sim 10^{11}$ m, which, in turn, means that deviations from Einstein gravity 
would occur at solar system scales; since these have not
been seen, we are constrained to take $d \geq 2$. For these values 
$R < 1$ mm, hence there is no conflict with known facts. It is also
perhaps worth mentioning that we would normally require $d <7$, since
that is the largest number allowed if the string theory is derivable
from M-theory, believed to be the fundamental theory of all interactions. 
In the ADD model
the smallness of Newton's constant is a direct consequence of the 
compactification-with-large-radius hypothesis
and hence there is no hierarchy problem in this theory\footnote{
A related problem, that of stabilization of the compactification scale,
exists, however; this has been discussed in Ref. \cite{ArDiMaRu}.}. 

In traditional KK theories, the mass-spectrum of nonzero KK modes 
arising from compactification of fields living in the bulk 
is driven to the Planck scale $M_{Pl}^{(4)}$. 
This problem is avoided in the ADD model 
by having the SM particles live on a `surface' with negligible width in the 
extra $d$ dimensions, which we identify with the 3-brane. The SM particles 
may then be thought of as excitations 
of open strings whose ends terminate on the brane; gravitons correspond
to excitations of closed strings propagating in the bulk. Thus,
the only interactions which go out of the 3-brane
into the bulk are gravitational ones.  We thus have a picture of a 
4-dimensional `surface' embedded in a ($4+d$)-dimensional space, where
SM fields live on the `surface', but gravitons can be radiated-off into 
the bulk. Noting that the SM fields are confined to the 3-brane, it is obvious
that the only new effects will be those due to exchange of gravitons
between particles on the 3-brane. To construct an effective theory
in 4 dimensions, gravity is quantized in the usual way, taking the 
weak-field limit, assuming that the underlying string theory takes care
of ultraviolet problems. The interactions of gravitons now follow 
from the ($4+d$)-dimensional Einstein equations in the compactification
limit. Feynman rules for this effective theory have been worked out in 
detail in Refs.~\cite{GiRaWe} and \cite{HaLyZh}. We use their prescriptions 
in our work. On the 3-brane, the couplings of the gravitons to the SM 
particles will be suppressed, as is well-known, by the Planck scale 
$M^{(4)}_{Pl} \simeq 1.2 \times 10^{19}$ GeV. This is offset, however, 
by the fact that, after compactification, the density of massive KK 
graviton states in the effective theory is very high, being, indeed
proportional to $M^{(4)}_{Pl}/M^{(4+d)}_{Pl}$. The Planck mass dependence
cancels out, therefore, leaving a suppression by the string scale 
$M_S \equiv M^{(4+d)}_{Pl} \simeq M_{EW}$. In the ADD theory, therefore, 
the tower of KK graviton states leads to effective interactions of 
{\it electroweak} strength. A further assumption made in our work ---
and in other phenomenological studies --- is that $Y$-particles, 
excitation modes of the 3-brane itself in the bulk --- are heavy and do not 
affect the processes under consideration. This corresponds to a static 
approximation for the brane. It is also relevant to mention that the
dilaton field associated with the graviton couples only to the trace of
the energy-momentum tensor, {\it i.e.} to the mass of the SM particles
at the vertex. For light fermions, as we have in the Drell-Yan process,
this means that the interactions of the dilaton can be safely neglected. 

Using these Feynman rules, it has been possible to explore a number of
different processes where the new interactions could cause observable
deviations from the SM. Only two new parameters enter the theory: one is 
the string scale $M_S \equiv M^{(4+d)}_{Pl}$. The other is a factor 
$\lambda$, of order unity and indeterminate sign, which arises when we 
sum over all possible KK modes of the graviton. As the amplitudes for 
virtual graviton exchange (with which we are concerned in this work) are
always proportional to $\lambda/M^4_S$, it is usual to absorb the
magnitude of $\lambda$ into $M_S$; this reduces the uncertainty to  
$\lambda = \pm 1$. Obviously this determines whether the graviton 
exchanges interfere constructively or destructively with the SM
interactions.

Remembering that the gravitons couple to any particle with a non-vanishing
energy-momentum tensor, it is possible to make a variety of phenomenoogical
studies of the new interactions and to test the workability of the ADD
model. Though the phenomenology of this model has not yet been fully explored,
several important results are already available in the literature. 
These can be classified into two types: those involving real KK graviton
production, and those involving virtual graviton exchange. A real KK mode of
the graviton will have interactions with matter suppressed by the Planck
scale $M^{(4)}_{Pl}$ and will therefore escape the detector. One can,
therefore, see signals with large missing momentum and energy if an observable
particle is produced in association with a KK graviton mode. However,
cross-sections for these depend explicitly on $d$, the number of extra
dimensions, and bounds derived from data reflect this dependence. Some of the processes 
examined so far include single-photon final states at $e^+e^-$ 
colliders\cite{GiRaWe,PeMiPe,ChKe} as well as hadron colliders\cite{GiRaWe}, monojet 
production at hadron colliders\cite{GiRaWe,PeMiPe}, 
two-photon processes at $e^+e^-$ colliders\cite{AtBaSo}, single-Z production
at $e^+e^-$ colliders\cite{ChKe}  
and the neutrino flux from the supernova SN1987A\cite{ADD3,CuPe}. 
Each process can be used 
to obtain a bound on the string scale $M_S$ for a given number $d$.
The most dramatic of these bounds is $M_S > 50$ TeV for $d = 2$ and it
comes from a study\cite{CuPe} of neutrinos from the supernova SN1987A. However,
this last bound drops to about a TeV as soon as we go to $d = 3$. Most of
the other processes lead to lower bounds of about 1--1.1 TeV on the 
string scale for $d = 2$, but these bounds become much weaker for
$d > 3$. 

Virtual (KK) graviton exchanges lead to extra contributions to processes
involving SM particles in the final state and can be observed as 
deviations in the cross-sections and distributions of these from the SM
prediction. After summation over all the KK modes of the graviton, the final result
is proportional to ${\rm sgn}(\lambda)/M^4_S$, with {\it practically 
no dependence on the number
of extra dimensions}\footnote{This is really because the density of graviton 
KK modes is approximated by a continuum, as a result of which mass degeneracies 
due to the number of extra dimensions are lost, at least to the leading order. 
In a sense, therefore, 
bounds from virtual graviton exchange are  more general.}.
Each process can be used to obtain a bound on the string scale $M_S$ for a 
given sign of $\lambda$. Some of the processes examined include Bhabha and 
M${\not{\rm o}}$ller 
scattering at $e^+e^-$ colliders\cite{GiRaWe,Rizzo2}, photon 
pair-production in $e^+e^-$\cite{GiRaWe,AgDe} and hadron colliders\cite{GiRaWe},  
fermion pair production in $\gamma\gamma$ colliders\cite{Rizzo2},
Drell-Yan production of dileptons\cite{Hewett}, dijet\cite{MaRaSr3} and 
top-quark\cite{MaRaSr1} pair production at hadron colliders, 
deep inelastic scattering at HERA\cite{MaRaSr2,Rizzo2}, massive 
vector-boson pair production in $e^+e^-$ collisions \cite{AgDe,Dicus} and pair 
production of scalars (Higgs bosons and squarks) at both $e^+e^-$ and 
$\gamma\gamma$ colliders\cite{Rizzo2}. Among the best
of these bounds is $M_S > 920~(980)$ GeV for $\lambda = +1(-1)$
which comes from
a study\cite{Hewett} of experimental data on Drell-Yan leptons at the Tevatron.
We make a more elaborate analysis if the same data in this work. 

The contributions to the Drell-Yan production of dileptons at hadron
colliders from graviton exchanges have been considered by 
Hewett\cite{Hewett}. Some of her findings relevant to the Tevatron are:
\begin{itemize}
\item There is very little difference between the cases $\lambda = +1$ 
and $\lambda = -1$ for the dilepton invariant mass distribution. 
\item The $\lambda = \pm 1$ cases differ, however, in the angular
distribution; therefore, widely differing forward-backward
asymmetries may be predicted. 
\item There are large deviations between the SM and the ADD model for
large invariant masses. 
\item The gluon-gluon contribution to the Drell-Yan process (see 
below) is much suppressed compared to the quark-initiated process. 
\item The bounds can increase to about 1.15~(1.35) TeV for 
$\lambda = +1(-1)$ in Run-II of the Tevatron. 
\end{itemize} 
We agree with most of these results at the generator level. However,
in the absence of published details about the angular distribution of 
dileptons observed by the CDF and D\O~Collaborations, we confine our
analysis to the invariant mass distributions only. Hence we do not
make a separate analysis for the two signs of $\lambda$. 

\begin{center}
\vspace*{-1.6in} 
% -----------------------------------------------------------------------
% This draws the Standard Model Feynman diagram for the Drell-Yan process
% -----------------------------------------------------------------------
\begin{picture}(400,220)(0,0)
\SetOffset(-15,0)
\SetWidth{2.0}
\SetScale{0.25}
% --------------------- Math ---------------------------------------------
\Line(-20,-60)(-20,370) 
% -------------------- Hadronic Junk -------------------------------------
\Line(50,0)(70,-50)
\Line(60,0)(80,-50)
\Line(70,0)(90,-50)
\Line(40,290)(70,340)
\Line(50,290)(80,340)
\Line(60,290)(90,340)
% -----------------Initial State Fermions --------------------------------
\ArrowLine(150,150)(50,0)
\ArrowLine(50,300)(150,150)
% ------------------Photon Propagator ------------------------------------
\Photon(150,150)(300,150){10}{5}
% -----------------Final State Fermions ----------------------------------
\ArrowLine(300,150)(400,0)
\ArrowLine(400,300)(300,150)
% --------------------- Vertices -----------------------------------------
\Vertex(140,150){10}
\Vertex(300,150){10}
% ----------------Blobs for Proton Breakup -------------------------------
\GOval(50,5)(15,10)(0){0.5}
\GOval(50,295)(15,10)(0){0.5}
% -----------------Proton and Antiproton ---------------------------------
\BBox(50,10)(-10,2)
\BBox(50,298)(-10,288)
% -------------------------Labels ----------------------------------------
\Text(25,75)[lb]{\Large $q$}
\Text(25,10)[lb]{\Large $\bar q$}
\Text(95,10)[lb]{\Large $\ell^-$}
\Text(85,75)[lb]{\Large $\ell^+$}
\Text(62,58)[c]{\large $\gamma^*,Z$}
% -----------------------------------------------------------------------
\Text(130,35)[c]{\huge\bf +}
\SetPFont{Times-Bold}{50}
\PText(220,-80)(0)[c]{(a)}
\PText(820,-80)(0)[c]{(b)}
\PText(1500,-80)(0)[c]{(c)}
% -----------------------------------------------------------------------
\end{picture}
% -----------------------------------------------------------------------

% -----------------------------------------------------------------------
% This draws the Kaluza-Klein quark-induced diagram for the Drell-Yan process
% -----------------------------------------------------------------------
\begin{picture}(400,220)(0,0)
\SetOffset(140,220)
\SetWidth{2.0}
\SetScale{0.25}
% --------------------- Math ---------------------------------------------
\Line(460,-60)(460,370) 
% -------------------- Hadronic Junk -------------------------------------
\Line(40,0)(70,-50)
\Line(50,0)(80,-50)
\Line(60,0)(90,-50)
\Line(40,290)(70,340)
\Line(50,290)(80,340)
\Line(60,290)(90,340)
% -----------------Initial State Fermions --------------------------------
\ArrowLine(150,150)(50,0)
\ArrowLine(50,300)(150,150)
% ------------------Graviton Propagator ------------------------------------
\ZigZag(150,150)(300,150){15}{15}
% -----------------Final State Fermions ----------------------------------
\ArrowLine(300,150)(400,0)
\ArrowLine(400,300)(300,150)
% --------------------- Vertices -----------------------------------------
\Vertex(140,150){10}
\Vertex(300,150){10}
% ----------------Blobs for Proton Breakup -------------------------------
\GOval(50,5)(15,10)(0){0.5}
\GOval(50,295)(15,10)(0){0.5}
% -----------------Proton and Antiproton ---------------------------------
\BBox(50,10)(-10,2)
\BBox(50,298)(-10,288)
% -------------------------Labels ----------------------------------------
\Text(35,75)[lb]{\Large $q$}
\Text(35,10)[lb]{\Large $\bar q$}
\Text(95,10)[lb]{\Large $\ell^-$}
\Text(90,75)[lb]{\Large $\ell^+$}
\Text(60,58)[c]{\large $G^*_n$}
\Text(122,90)[c]{\large 2}
\end{picture}
% -----------------------------------------------------------------------

% -----------------------------------------------------------------------
% This draws the Kaluza-Klein gluon-induced diagram for the Drell-Yan process
% -----------------------------------------------------------------------
\begin{picture}(400,20)(0,-200)
\SetOffset(300,45)
\SetWidth{2.0}
\SetScale{0.25}
% --------------------- Math ---------------------------------------------
\Line(-40,-60)(-40,370) 
\Line(460,-60)(460,370) 
% -------------------- Hadronic Junk -------------------------------------
\Line(40,0)(70,-50)
\Line(50,0)(80,-50)
\Line(60,0)(90,-50)
\Line(40,290)(70,340)
\Line(50,290)(80,340)
\Line(60,290)(90,340)
% -----------------Initial State Gluons-----------------------------------
\Gluon(150,150)(50,0){-12}{5}
\Gluon(50,300)(150,150){-12}{5}
% ------------------Graviton Propagator ------------------------------------
\ZigZag(150,150)(300,150){15}{15}
% -----------------Final State Fermions ----------------------------------
\ArrowLine(300,150)(400,0)
\ArrowLine(400,300)(300,150)
% --------------------- Vertices -----------------------------------------
\Vertex(140,150){10}
\Vertex(300,150){10}
% ----------------Blobs for Proton Breakup -------------------------------
\GOval(50,5)(15,10)(0){0.5}
\GOval(50,295)(15,10)(0){0.5}
% -----------------Proton and Antiproton ---------------------------------
\BBox(50,10)(-10,2)
\BBox(50,298)(-10,288)
% -------------------------Labels ----------------------------------------
\Text(-30,35)[c]{\huge\bf +}
\Text(35,75)[lb]{\Large $g$}
\Text(35,10)[lb]{\Large $g$}
\Text(95,10)[lb]{\Large $\ell^-$}
\Text(90,75)[lb]{\Large $\ell^+$}
\Text(60,58)[c]{\large $G^*_n$}
\Text(122,90)[c]{\large 2}
\end{picture}
% -----------------------------------------------------------------------
\end{center}
\vspace*{-3.0in} 

\noindent
{\bf Figure 1}. 
{\footnotesize\it Feynman diagrams for the contribution
to the Drell-Yan process from {\em (a)} the Standard Model and {\em (b,c)}
exchange of a Kaluza-Klein graviton.} \\

\def\hs{\hat{s}}
\def\hT{\hat{t}}
\def\hu{\hat{u}}
\def\sw{\sin\theta_W}
\def\cw{\cos\theta_W}
\def\s2w{\sin^2\theta_W}
\def\c2w{\cos^2\theta_W}

The Drell-Yan cross-section, including the effects of Kaluza-Klein
graviton exchanges, is given by the above Feynman diagrams. The
Standard Model diagrams ($a$) involving exchange of a photon or
a $Z$-boson in the $s$-channel, interfere with the diagram with
$s$-channel exchange of a Kaluza-Klein graviton ($b$), while the
diagram ($c$) has no Standard Model analogue.

Evaluating these leads to the result
\begin{eqnarray}
&& \sigma_{DY} (p\bar p \rightarrow \ell^+ \ell^-) =
\int dx_1 dx_2~f_{g/p}(x_1)~f_{g/\bar p}(x_2)~
\hat{\sigma}(gg \rightarrow \ell^+ \ell^-) \\ && +
\sum_{q = u,d,s} \int dx_1 dx_2~[f_{q/p}(x_1)~f_{\bar q/\bar p}(x_2)~
+f_{\bar q/p}(x_1)~f_{q/\bar p}(x_2)]~
\hat{\sigma}(q\bar q \rightarrow \ell^+ \ell^-)~, \nonumber
\end{eqnarray}
where $f_{a/b}(x)$ denotes the flux of a parton $a$ in a beam of particles
$b$, 
\begin{equation}
\hat{\sigma}(q\bar q ~{\rm or}~gg \rightarrow \ell^+ \ell^-)
= \frac{1}{16\pi \hs^2}~
\overline{|{\cal M}(q\bar q ~{\rm or}~gg \rightarrow \ell^+ \ell^-)|^2}~,
\end{equation}
and $\overline{|{\cal M}|^2}$ represents the squared Feynman amplitude
summed over final spins and averaged over initial spins and colours.

Evaluation of the Feynman diagrams gives, for the gluon-induced process
(which has no Standard Model analogue):
\begin{equation}
\overline{|{\cal M}(gg \rightarrow \ell^+ \ell^-)|^2}
= \bigg( \frac{\pi}{2 M_S^4} \bigg)^2
\bigg[ \hs^4 + 2 \hT \hu~(\hT - \hu)^2 \bigg]~,
\end{equation}
when all the graviton Kaluza-Klein modes have been summed over.

Evaluation of the Feynman diagrams gives for the quark-induced process
(including interference terms):
\begin{equation}
\overline{|{\cal M}(q\bar q \rightarrow \ell^+ \ell^-)|^2}
= T^{q\bar q}_{SM} + T^{q\bar q}_{KK}~,
\end{equation}
where the Standard Model contribution is given below. We adopt the
convention that $T_a$ denotes the contribution from exchange of
a particle $a$ and $T_{ab}$ denotes the interference term between
diagrams with exchange of $a$ and $b$ respectively. With these, we
get
\begin{eqnarray}
T^{q\bar q}_{SM}
& = & T^{q\bar q}_\gamma + T^{q\bar q}_Z + T^{q\bar q}_{Z\gamma}~;
\\
T^{q\bar q}_\gamma
& = & \frac{32}{3} (\pi\alpha Q_q)^2~\bigg[\frac{\hT^2 + \hu^2}{\hs^2} \bigg]~,
\nonumber \\
T^{q\bar q}_Z
& = & \frac{1}{3} \bigg(\frac{\pi\alpha}{4\s2w\c2w} \bigg)^2
|D_Z(\hs)|^2 \bigg[ (L^2_\ell L_q^2 + R^2_\ell R^2_q )~\hT^2
                  + (L^2_\ell R_q^2 + R^2_\ell L^2_q )~\hu^2 \bigg]~,
\nonumber \\
T^{q\bar q}_{Z\gamma}
& = & - \frac{2}{3} Q_q \bigg(\frac{\pi\alpha}{\sw\cw}\bigg)^2
|D_Z(\hs)|^2~\bigg( 1 - \frac{M^2_Z}{\hs} \bigg)
\nonumber \\
&& \times ~\bigg[ (L_\ell L_q + R_\ell R_q )~\hT^2
      +~(L_\ell R_q + R_\ell L_q )~\hu^2 \bigg]~,
\nonumber
\end{eqnarray}
defining
\begin{equation}
D_Z(\hs) = \big[ \hs - M_Z^2 + i M_Z \Gamma_Z \big]^{-1}
\end{equation}
and
\begin{displaymath}
L_\ell   =  4 \sin^2 \theta_W - 2~, \qquad
R_\ell   =  4 \sin^2 \theta_W~,
\end{displaymath}
\begin{displaymath}
L_q      =  4 (T_{3q} - Q_q \sin^2 \theta_W )~, \qquad
R_q      =  - 4 Q_q \sin^2 \theta_W ~,
\end{displaymath}
for the couplings. The non-Standard part, using the same convention,
is given by
\begin{eqnarray}
T^{q\bar q}_{KK}
& = & T^{q\bar q}_G + T^{q\bar q}_{G\gamma} + T^{q\bar q}_{GZ}~,
\\
T^{q\bar q}_G
& = & \frac{\lambda^2}{3} \bigg( \frac{\pi}{2 M^4_S} \bigg)^2
\bigg[ \hs^4 - 4 \hs^2~(\hT - \hu)^2
+ (\hT - \hu)^2 ~(5\hT^2 - 6\hT\hu + 5\hu^2) \bigg]~,
\nonumber \\
T^{q\bar q}_{G\gamma}
& = & -\frac{4}{3} \lambda Q_q \frac{\pi^2 \alpha}{M^4_S}
\bigg( \frac{\hT - \hu}{\hs} \bigg)
\bigg[ \hs^2 - 2~(\hT^2 + \hu^2) - (\hT - \hu)^2 \bigg]~,
\nonumber \\
T^{q\bar q}_{GZ}
& = & - \lambda \frac{\alpha}{3} \bigg( \frac{\pi}{2\sw\cw M^2_S} \bigg)^2
(\hs - M^2_Z)~ |D_Z(\hs)|^2
\nonumber \\
&& \times \bigg[ (L_\ell L_q + R_\ell R_q)~\hT^2~(\hT - 3 \hu)
     -~(L_\ell R_q + R_\ell L_q)~\hu^2~(\hu - 3 \hT) \bigg]~,
\nonumber
\end{eqnarray}
when, as before, all the Kaluza-Klein modes have been summed over.

The above formulae represent the lowest order (LO) calculation
in perturbation theory. The calculation of higher-order effects,
especially next-to-leading order (NLO) and next-to-NLO (NNLO)
QCD corrections has been done in detail\cite{vanN} for the SM process
represented by $T^{q\bar q}_{SM}$. No corresponding calculations
have been attempted as yet for the KK parts, $T^{q\bar q}_{KK}$
and ${\cal M}(gg \rightarrow \ell^+ \ell^-)$. In the absence
of such a calculation, we make the assumption that the change
in the LO cross-section due to QCD corrections --- the `K-factor' 
---is identical for 
the SM and KK parts. Our results are, therefore, correct only
within this approximation\footnote{This places our results on an
equal footing with a large number of experimental bounds on new physics 
scenarios, such as those involving quark and lepton 
compositeness\cite{D0_PRL}, 
for which the QCD corrections are not available.}. However, we do not 
expect a proper calculation of NLO effects to make a drastic change in 
our rough-and-ready results,
because the dominant contribution to dilepton production at the
Tevatron comes from quark-induced processes. Since the SM and
KK results both arise from colour-singlet exchange, the actual
`K-factor' is likely to be rather similar in both cases. For gluons,
this is not true, but the gluon-induced process makes only a minor
contribution at Tevatron energies. 

In keeping with this philosophy, therefore, we have extracted,
for each value of the dilepton invariant mass $M \equiv M_{\ell^+\ell^-}$,
a `K-factor' by taking the ratio of the LO SM cross-section calculated
using the above formulae with that calculated using the full NNLO
calculation of Ref.~\cite{vanN}. This set of ratios is then used to 
scale the entire differential cross-section when the KK effects
are included. It is worth pointing out that this procedure also
takes care of the leading effects arising from initial-state radiation.
Finally, it is relevant to mention that we have used the CTEQ-4M set of
structure functions\cite{CTEQ4M} to calculate the initial state parton luminosities.  

We now describe our analysis in some detail. The D\O~Collaboration has
presented\cite{D0_PRL} the $e^+e^-$ 
invariant mass distribution in 9 bins starting from 
120 GeV till 1 TeV using the di-electron data collected with 120
pb$^{-1}$ of luminosity. The cuts relevant for the cross-section 
calculation are given below. No distinction is made between the electron
and the positron. 
\begin{itemize}
\item The transverse momentum of both the isolated electrons must satisfy 
$p_T > 25$ GeV.
\item The electrons are called CC (for Central Calorimeter) if
they satisfy $|\eta| <1.1$, $\eta$ being the pseudorapidity;
they are called EC (for End Cap) if they satisfy 
$1.5 < |\eta| < 2.5$. 
\end{itemize}
Only those events are considered in which there is
at least one CC electron, while the other can be CC or EC. 
The acceptances described above are taken into account while estimating
our Monte Carlo cross-sections. These cross-sections need to be further 
convoluted with efficiencies\cite{D0_PRL} which are ($74.1\pm0.6$)\% 
when both electrons are CC and ($52.6\pm1.0$)\% when one of them is EC. 
Multiplying by the luminosity now gives us a prediction for the number
of di-electron events expected in each mass bin, which is then compared with
the D\O~data.

The CDF Collaboration has presented\cite{CDF_PRD} results for  
dimuon samples, using 107 pb$^{-1}$ of data. The relevant cuts are given below.
\begin{itemize}
\item The reconstructed rapidity $y$ of the virtual $s$-channel state
(`boson rapidity') is required to satisfy $|y|<1$ for all events.
\item Both muons are required to satisfy $|\eta|<1$, which confines
the analysis to the central region. 
\item A back-to-back cut $|\eta_1 + \eta_2| \geq 0.2$ is
imposed: this gets rid of cosmic ray backgrounds.
\item Both muons are required to satisfy a `loose' transverse momentum
cut of $p_T > 17$ GeV and at least one is required to satisfy
a `tight' cut of $p_T > 20$ GeV. 
\end{itemize}
These cuts are applied in our Monte Carlo generator to estimate the
cross-section times acceptance for the 6 mass bins
in the range 120 GeV to 500 GeV presented in Ref.~\cite{CDF_PRD} (Table X).
These are convoluted with the experimental efficiencies (Table VI of 
Ref.~\cite{CDF_PRD}).
We then obtain an additional 
correction factor for each mass bin by normalising our
SM expectation to the numbers given in Ref.~\cite{CDF_PRD}. 
This may be expected to take care of the effect of other detector-specific
cuts like triggers, etc. Finally,
we use this correction factor along with our generator-level acceptance
and the experimental efficiencies 
to estimate the number of events in each mass bin for various values
of $M_S$. The choice of only 6 mass bins in the range 120 GeV to 500 GeV 
is because
the ADD model predicts wider deviations from the SM in the higher mass
bins (see Fig.~2). 
We also take note of the fact that no events are seen at CDF in the mass 
bin 500 GeV to 1 TeV. 

% ------------------------------------------------------------------
\begin{figure}[htb]
\begin{center}
\vspace*{3.2in}
      \relax\noindent\hskip -4.4in\relax{\includegraphics{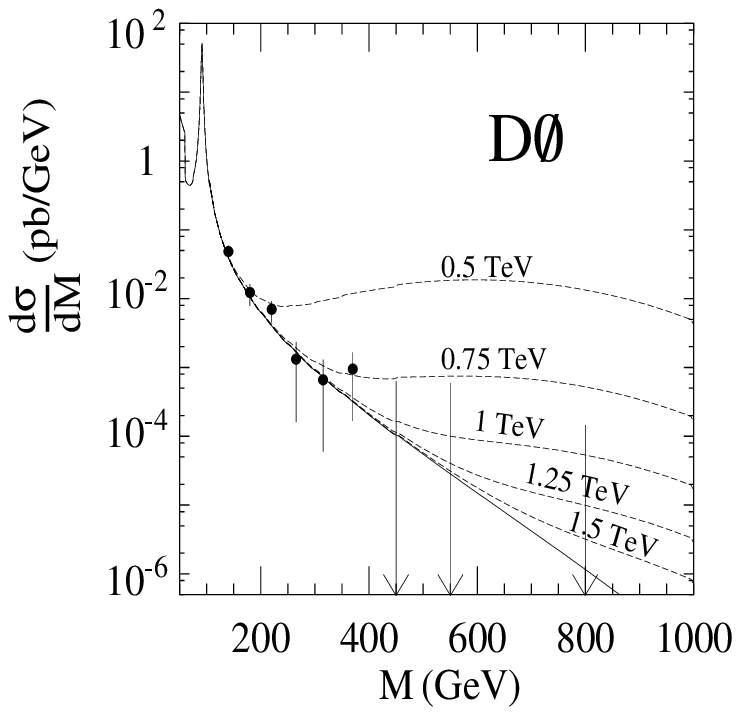}}
                     \hskip  2.8in\relax{\includegraphics{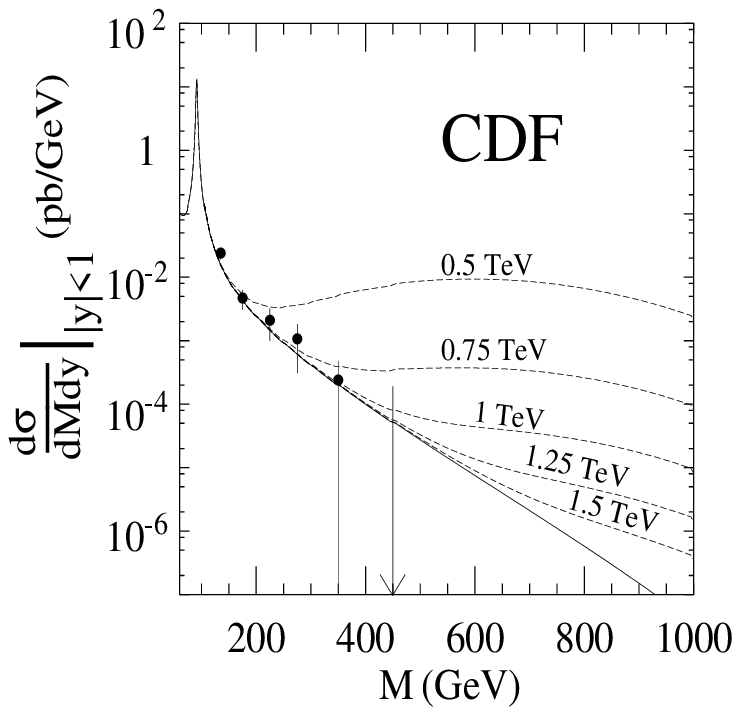}}
\end{center}
\end{figure}
\vspace*{-1.0in}
\noindent {\bf Figure 2}. 
{\footnotesize\it Illustrating the effects of TeV scale quantum gravity
on the invariant mass distributions of dileptons seen at the Tevatron
by the D\O~and CDF Collaborations respectively. Solid lines show the SM 
prediction; dashed lines show the predictions of the ADD model for
marked values of $M_S$. }
% ------------------------------------------------------------------
\vskip 5pt

In Fig.~2 we show the differential cross-section as a
function of the invariant mass $M$ of the dilepton, compared to the
D\O~and CDF data. We have set $\lambda = +1$, but 
$\lambda = -1$ will not make a discernible change in the figure. Solid lines 
show the SM
prediction; dashed lines show the predictions of the ADD model for
$M_S =$ 0.5, 0.75, 1, 1.25 and 1.5 TeV respectively. The data points
correspond to those used in our analysis and do not represent the full
set of available points. Error bars are presented at 68\% confidence level 
(C.L.) if there
are events in the relevant mass bin and a 95\% C.L. upper bound if
there are no events in the relevant mass bin.
The D\O~numbers correspond to a differential cross-section
$d\sigma/dM$: this is obtained by dividing the cross-section
(modulo cuts) in the mass bin by the width of that bin.
The CDF numbers correspond to a double differential cross-section
$d^2\sigma/dMdy$ in
both $M$ and $y$: this is obtained by dividing the cross-section
(modulo cuts) in the mass bin by the width of that bin as well
as by a factor $\Delta y = 2$.

As is apparent from the figure, the string scale cannot be anywhere
near 500 GeV, since that would show extreme deviations from the observed
data. This is just one of the arguments which tells us that quantum
gravity effects must lie at scales of a TeV or more. On the other hand,
as $M_S$ approaches 1 TeV, the differentiation
between signal and background is less striking. This is partly because the 
deviations
arise only in the high mass bins, where no events are expected with
the current luminosities. 

The actual limits on the string scale $M_S$ are calculated using a Bayesian 
analysis of the shape of the mass distribution of events. For a value $M_S$
of the string scale, the expected number of events in the $k^{\rm th}$ mass 
bin can be written as:
\begin{eqnarray} 
 N^k(M_S) = b_k + {\cal L}~{\epsilon_k}~{\sigma^k(M_S)} 
\end{eqnarray}
where ${\cal L}$ is the data luminosity, $b_k$ is the expected background, 
${\epsilon_k}$ is the dilepton detection efficiency and 
$\sigma^k(M_S)$ is the expected dielectron cross section with inclusion
of the effect due to large extra dimension. 

The posterior probability density for the string scale to be $M_S$, 
given the observed data distribution $(D)$, is given by
\begin{eqnarray}
P({M_S}{\mid}{D})=
\frac{1}{A}
\int db~{d\epsilon}~d{\cal L}
\prod_{k=1}^n 
\left[ 
\frac { {e^{-{N^k(M_S)}}}{N^k(M_S)}^{N_{0}^k} } {{N_{0}^k}!}
\right] 
P({b}, {\cal L}{\epsilon})~P({M_S}). 
\label{post_prob}
\end{eqnarray} 
In the above equation the term in square brackets is the likelihood for the 
data distribution to be from a model with string scale $M_S$. The prior probability 
$P({b},{\cal L}{\epsilon})$ is taken to be a product of independent Gaussian 
distributions in $b$, ${\cal L}$ and $\epsilon$, with the measured value in each bin 
defining the mean and the uncertainity defining the width. The overall factor
$1/A$ is just a normalisation. 
Since the excess cross-sections due to graviton exchanges are combinations of 
direct terms proportional to 1/${M_S^8}$ and interference terms proportional
to  1/${M_S^4}$, we consider a prior distribution $P(M_S)$ uniform in 
($a$)~1/${M_S^4}$ and ($b$)~1/${M_S^8}$ separately. The limit on the string scale from 
a prior uniform in 1/${M_S^8}$ represents a conservative estimate; using a prior
uniform in 1/${M_S^4}$ provides more stringent limits. 
From the above posterior probability, the cumulative probability  
= $\int_{M_S}^{\infty} P({M_S^\prime}{\mid}{D}) d{M_S^\prime}$ 
can be calculated. 
The $M_S$ value at which the cumulative probability equals 0.95 is, then, the 95\%
C.L. limit. We also combine the data using the simple expedient of treating 
the CDF probability as a prior for the D0 analysis (and vice versa). 

% ------------------------------------------------------------------
\begin{figure}[htb]
\begin{center}
\vspace*{3.2in}
      \relax\noindent\hskip -2.9in\relax{\includegraphics{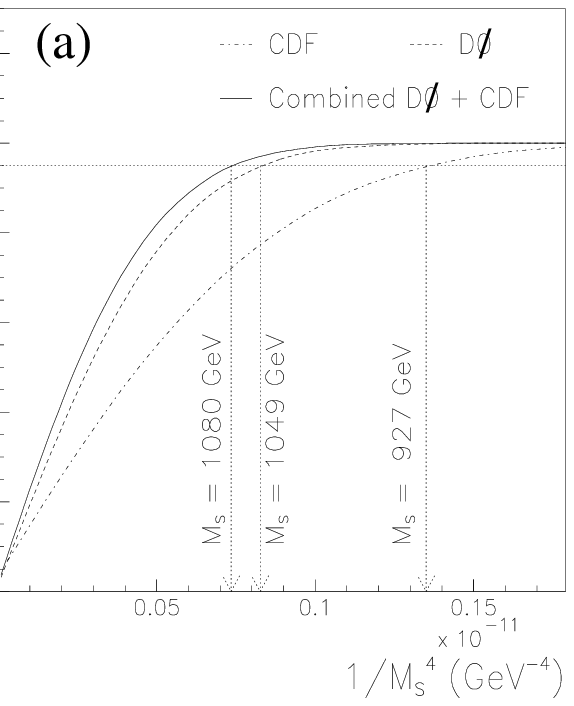}}
                     \hskip  2.0in\relax{\includegraphics{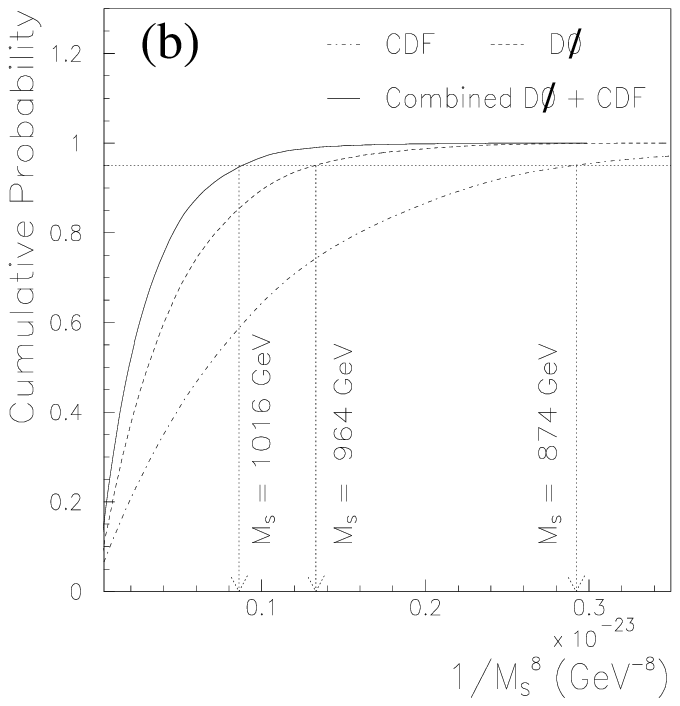}}
\end{center}
\end{figure}
\vspace*{-0.9in}
\noindent {\bf Figure 3}.
{\footnotesize\it Showing the (cumulative) posterior probability for
the ADD model with different values of the string scale $M_S$ 
assuming a prior probability which is {\em (a)} uniform in $1/M_S^4$
and {\em (b)} uniform in $1/M_S^8$. }
% ------------------------------------------------------------------
\vskip 5pt

In Fig.~3, we have plotted the cumulative posterior probability for
the ADD model with string scale $M_S$ as a function
of ($a$)~$1/M_S^4$ and ($b$)~$1/M_S^8$. The dashed (dash-dot) lines
indicate the results of 
considering the D\O~(CDF) data alone, while the solid lines show the 
result of a combined fit. A glance at the figure will show that
the horizontal (dotted) lines correspond to 95\% C.L. limits,
while the fact that the curve saturates for higher values of
$1/M_S^{(4,8)}$ shows that the SM values ($M_S \rightarrow \infty$)
constitute the best hypothesis to fit the data. In a more 
quantitative idiom, we may 
interpret the vertical (dotted) lines as lower bounds on 
the string scale $M_S$. It is clear that the bound of 927 GeV assuming
a prior probability of 1/$M_S^4$, and using the CDF data,
is consistent with that reported in Ref.~\cite{Hewett},
while the value 874 GeV assuming a prior probability of 1/$M_S^8$ 
represents a more conservative estimate. 
The D\O~data provide an improvement\footnote{This is for two reasons:
($a$) the published results from D\O~use a slightly higher integrated luminosity and
($b$) the D\O~Collaboration presents more data in the higher mass bins ---
where most of the deviations lie --- than the CDF Collaboration.} 
in the bound by about 100 GeV in both cases. 
Since the cross-section varies principally 
as $1/M_S^8$ in the region around $M_S = 1$ TeV (this is reflected
in the fact that it depends very weakly on the sign of $\lambda$), 
this corresponds
to an increase in the sensitivity by a factor of about 2.6. Combining the data
increases the sensitivity by another factor of about 1.2, which takes
the bound to 1080 (1016) GeV, depending on the choice of prior probability. 
Increasing the energy to 2 TeV
and the luminosity to 2 fb$^{-1}$, which may be expected with the
commissioning of the Main Injector in Run-II, improves the bounds by
a further 200--300 GeV; this corresponds to an improvement in the sensitivity
by a factor close to 4.  

To conclude then, we have used published dilepton data from the D\O~and CDF 
Collaborations to put bounds on the string scale $M_S$. This is 
the fundamental scale of the ADD model, which envisages large compact
dimensions in addition to the known (noncompact) ones and predicts strong 
quantum gravity effects at TeV scales.
Only the invariant mass distribution has been used and not the 
angular distribution. The latter might show some sensitivity to
the sign of $\lambda$. For the current analysis, however, there is
hardly any such sensitivity. Our result is also independent of the number
of extra dimensions $d$. 
We obtain a bound on $M_S$ of 900 GeV (900 GeV -- 1 TeV) using 
CDF (D\O) data alone
and a bound of around 1.0 -- 1.1 TeV using the combined data from both
experiments. This is one of the most stringent bound obtained from collider
studies at the present time and is likely to be improved
(to about 1.3 TeV) in Run-II of the Tevatron. 

The authors would like to thank Ashoke Sen and K.~Sridhar for reading 
the manuscript. We also acknowledge fruitful discussions with 
Dilip K. Ghosh, Supriya Jain, Gautam Mandal, Prakash Mathews and Caramine 
Pagliarone.

\end{document}